\documentclass[a4paper,fleqn,usenatbib]{mnras}

\usepackage[T1]{fontenc}
\usepackage{ae,aecompl}

\hypersetup{breaklinks=true}

\usepackage{graphicx}	
\usepackage{amsmath}	
\usepackage{amssymb}	

\newcommand{\pder}[2]{\frac{\partial#1}{\partial#2}}

\newcommand{\dbtbt}{\delta B^2/B^2}

\title[Early cross-field propagation of particles]{Early
  propagation of energetic particles across the mean field in
  turbulent plasmas}

\author[T. Laitinen et al.]{
T. Laitinen\thanks{E-mail: tlmlaitinen@uclan.ac.uk},
S. Dalla
and D. Marriott
\\
Jeremiah Horrocks Institute, University of Central Lancashire, Preston PR1 2HE, UK
}

\date{Accepted XXX. Received YYY; in original form ZZZ}

\pubyear{2016}

\begin{document}
\label{firstpage}
\pagerange{\pageref{firstpage}--\pageref{lastpage}}
\maketitle

\begin{abstract} 
  Propagation of energetic particles across the mean field direction
  in turbulent magnetic fields is often described as spatial
  diffusion. Recently, it has been suggested that initially the
  particles propagate systematically along meandering field lines, and
  only later reach the time-asymptotic diffusive cross-field
  propagation. In this paper, we analyse cross-field propagation of
  1--100~MeV protons in composite 2D-slab turbulence superposed on a
  constant background magnetic field, using full-orbit particle
  simulations, to study the non-diffusive phase of particle
  propagation with a wide range of turbulence parameters. We show that
  the early-time non-diffusive propagation of the particles is
  consistent with particle propagation along turbulently meandering
  field lines. This results in a wide cross-field extent of the
  particles already at the initial arrival of particles to a given
  distance along the mean field direction, unlike when using spatial
  diffusion particle transport models. The cross-field extent of the
  particle distribution remains constant for up to tens of hours in
  turbulence environment consistent with the inner heliosphere during
  solar energetic particle events. Subsequently, the particles escape
  from their initial meandering field lines, and
  the particle propagation across the mean field reaches
  time-asymptotic diffusion. Our analysis shows that in order to
  understand solar energetic particle event origins, particle
  transport modelling must include non-diffusive particle propagation
  along meandering field lines.
\end{abstract}




\section{Introduction}

Understanding the propagation of energetic particles in turbulent
plasmas is the key for understanding the acceleration of solar
energetic particles (SEPs), and their relation to the complex phenomena
during solar eruptions. The propagation of these charged particles is
affected by the heliospheric electric and magnetic fields. The
particles are guided by the Parker spiral field, and experience
guiding centre drifts across the field \citep[e.g.][]{Marsh2013}. The
turbulent fluctuations in the magnetic field, on the other hand, bring
a stochastic element to the propagation of the particles, often
modelled as diffusion along and across the mean field direction
\citep{Parker1965}.

Energetic particle propagation across the mean magnetic field in
turbulent plasmas has been considered as mainly the effect of
particles following the meandering field lines
\citep[e.g.][]{Jokipii1966}. Current approaches aiming to quantify
this effect take into account scattering of the particles along the
field lines \citep{Matthaeus2003,Shalchi2010a}. Also the decoupling of
the particles from the field lines has been considered
\citep{Fraschetti2011perptimetheory,RuffoloEa2012}. These theoretical
approaches work towards a time-asymptotic description of particle
cross-field transport as a spatial diffusion process. Cross-field
diffusion description has recently been applied also in modelling the
SEP propagation in the heliosphere \citep{Zhang2009,Droge2010, He2011,
  Tautz2011, Giajok2012, Qin2013,Strauss2017perpel}.

However, \citet{LaEa2013b} noted recently that early in the
propagation history, particles propagate systematically along
meandering field lines, spreading efficiently across the mean field
direction \citep[see also][]{Tooprakai2016}. This spreading is
non-diffusive in nature, and only at later times the particles
decouple sufficiently from their meandering field lines, resulting in
propagation that can be described as diffusion across the mean
magnetic field. Using full-orbit particle simulations in a cartesian
geometry, \citet{LaEa2013b} concluded that the temporal and spatial
evolution of impulsively-injected 10 MeV protons, as recorded 1 AU
from the injection region, remained inconsistent with diffusion
description for $\sim 20$~hours after the particle injection. Thus,
the non-diffusive early propagation may be very significant to the
propagation of the SEPs from the Sun to the
Earth. \citet{LaEa2016parkermeand} showed that the wide SEP events
observed with multiple spacecraft at different heliographic longitudes
at 1~AU
\citep[e.g.][]{Lario2013,Dresing2012,Wiedenbeck2013,Dresing2014,Cohen2014,Richardson2014}
could be explained using this approach with particle transport
parameters consistent with the interplanetary turbulence properties
already with a narrow source at the Sun.

The initial study by \citet{LaEa2013b} only addressed one set of
particle and turbulence parameters, and did not explore the parameter
space of particle and turbulence further to identify properties that
may influence the initial non-diffusive particle propagation phase. In
this work, we will study the initial non-diffusive phase and the
asymptotic diffusive phase in more detail, varying both the particle
and turbulence parameters. We will study the nature of the transition
from the initial to the asymptotic phase guided by the findings of
\citet{LaDa2017decouple}, who used a novel method to quantify how the
particles are displaced from the meandering field lines and discovered
that initially the particles are tied to their meandering field lines
well, and decouple only at later stages. We explain the particle
simulations and the analysis methods in Section~\ref{sec:model} and
Appendix~\ref{sec:width-distr-simple}, present our results and discuss
their relevance in Sections~\ref{sec:results}
and~\ref{sec:discussion}, and draw our conclusions in
Section~\ref{sec:conclusions}.

\section{Model}\label{sec:model}

The particles are simulated in magnetic field given by 
\begin{equation}\label{eq:turbfield}
\mathbf{B}(x,y,z)=B \hat{\mathbf{z}}+\delta \mathbf{B}(x,y,z),
\end{equation}
where $B$ is a constant background field, along the $z$-axis, for
which we use the value 5~nT, consistent with magnetic field at
1~AU. The fluctuating component $\delta \mathbf{B}(x,y,z)$ consists of
Fourier modes, and is constructed using the method presented in
\citet{GiaJok1999}, and fulfills $\nabla\cdot\mathbf{B}=0$.
We use a composite model, where turbulence is composed of slab and 2D
components, at power ratio 20\%:80\%. The turbulence is
  axisymmetric, with axissymmetric distribution of polarisation and
  wave vector directions. The turbulence amplitude is parametrised by
using the relative amplitude $\dbtbt$, where $\delta B^2$ is the
variance of the fluctuations. The turbulence spectrum for the 2D
component follows the Kolmogorov scaling, whereas for the slab
component we use spectral indices $q_\parallel=5/3$ and $1$.

The full-orbit particle simulations follow the same approach as
\citet{LaEa2012}. We start the particles in a large volume, to reduce
the effect of local magnetic field structures, but within the analysis
the particle location at time $t$ is determined relative to each
particle's initial position. The particles are injected into the
simulation as a beam, with pitch angle cosine $\mu=1$, and simulated
for 60 hours.

\begin{figure}
  \includegraphics[
  width=\columnwidth]{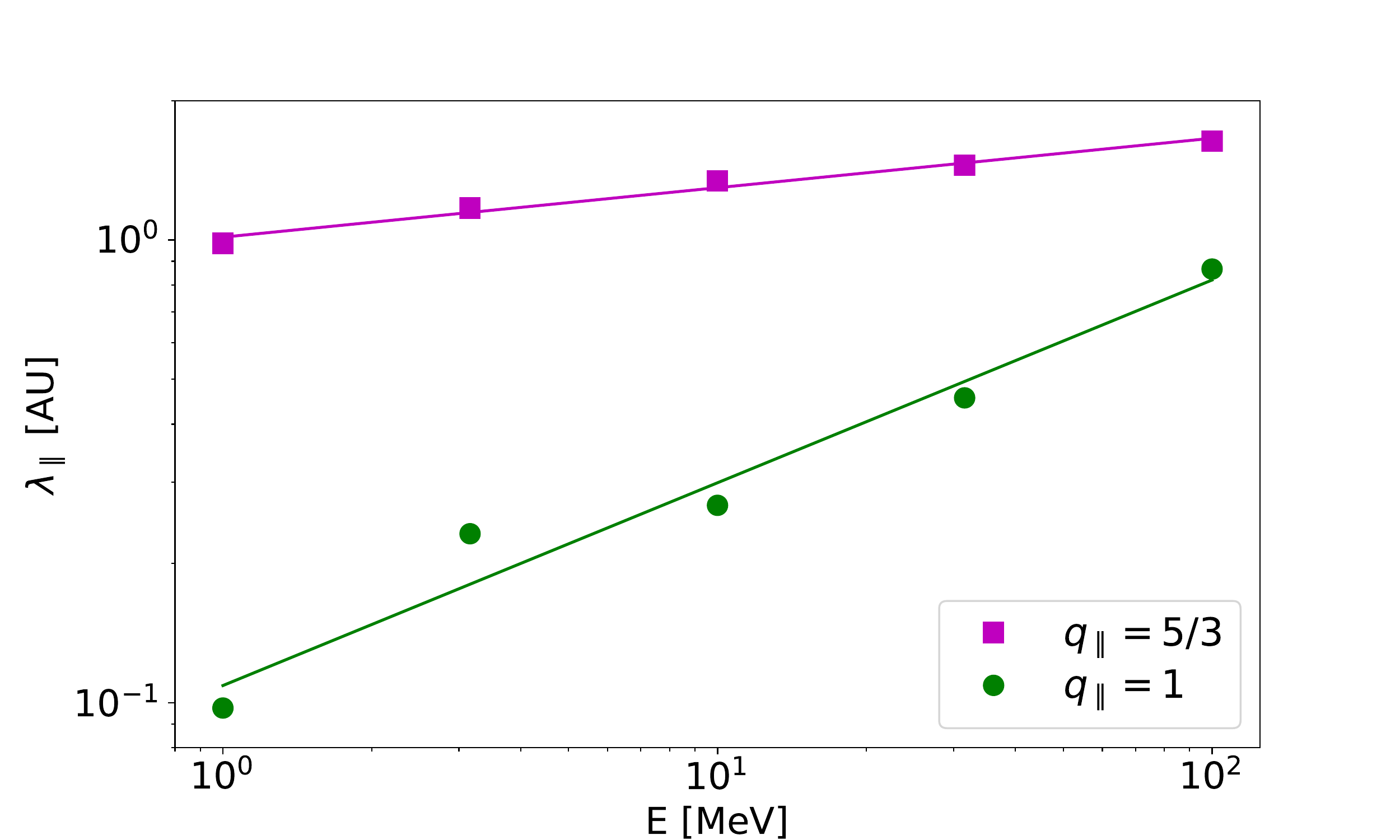}\\   
    \caption{The parallel mean free path of protons at energies
      1--100~MeV for turbulence with $\dbtbt=0.1$. The curves show the power-law fits to the mean free paths, with trends $E^{0.11}$ and $E^{0.44}$ for $q_\parallel=5/3$ and $1$, respectively.
      \label{fig:lambdaparE}}
\end{figure}

The slab spectral index affects the particles' parallel scattering
mean free path, which from quasilinear theory \citep{Jokipii1966}
varies as $\lambda_\parallel\propto R^{2-q_\parallel}$, where $R$ is
the particle rigidity. In addition, the turbulence energy for the
spectra with the two slab spectral indices, $q_\parallel=5/3$ and $1$,
results in a different scattering power at the resonant scales of the
particles simulated in this study. We show $\lambda_\parallel$ as
determined from the particle simulations in
Figure~\ref{fig:lambdaparE}, for the two spectral indices and
$\dbtbt=0.1$. The trends of $\lambda_\parallel$ as function of proton
energy for the two slab spectral indices, depicted with the fitted
power law curves, with $E^{0.11}$ and $E^{0.44}$ for $q_\parallel=5/3$
and $1$, respectively, are consistent with the quasilinear theory
result. The mean free paths presented in Figure~\ref{fig:lambdaparE}
are consistent with those obtained using interplanetary turbulence
properties
\citep[e.g.][]{Pei2011difftens,LaEa2016parkermeand,Strauss2017perpel} and SEP
observation analysis \citep[e.g.][]{Palmer1982,Torsti2004}.

We analyse the cross-field extent of the particle distribution as a
function of the distance along the mean field direction, using
\begin{equation}
  \label{eq:sigma}
  \sigma_\perp^2(z,t)=\left<(x(z,t)-\left<x(z,t)\right>)^2\right>
\end{equation}
where $\left<\right>$ represents the ensemble average of particles.
The use of $x$ instead of $y$ is arbitrary, and due to the
  axissymmetry of the turbulence has no effect on the obtained
  values. Also deviations in $r_\perp=\sqrt{x^2+y^2}$ could be
  considered, however it would complicate comparing our results with
  other work which typically consider cartesian deviations and
  diffusion coefficients. We note also that Equation~(\ref{eq:sigma})
  calculates the deviations with respect to the mean,
  $\left<x(z,t)\right>$, in order to eliminate the effects due to the
  potential asymmetries of the distributions (see
  Figures~\ref{fig:econtours} and~\ref{fig:bcontours}) that arise from
  finite range of fluctuation scales used in the simulations.

The form of the
  Equation~(\ref{eq:sigma}) differs from the conventional definitions
  in that $\sigma_\perp^2(z,t)$ is defined for particles at a given
  field-parallel distance $z$ at time $t$, instead of for all
  particles in the simulation. This choice is motivated by
observations: we do not observe the full 3D distribution of the
particles, but rather sample the particle distribution in fixed points
in space. Recent observations of SEPs by the STEREO, SOHO and ACE
spacecraft have provided us with a view of the longitudinal extent of
SEP events at 1~AU from the Sun. Our definition of $\sigma_\perp^2$
aims to provide comparison of simulated particle tranport with these
measurements.  

\section{Results}\label{sec:results}

\begin{figure*}
  \includegraphics[trim=5mm 20mm 5mm 5mm, clip=true,
  width=2\columnwidth]{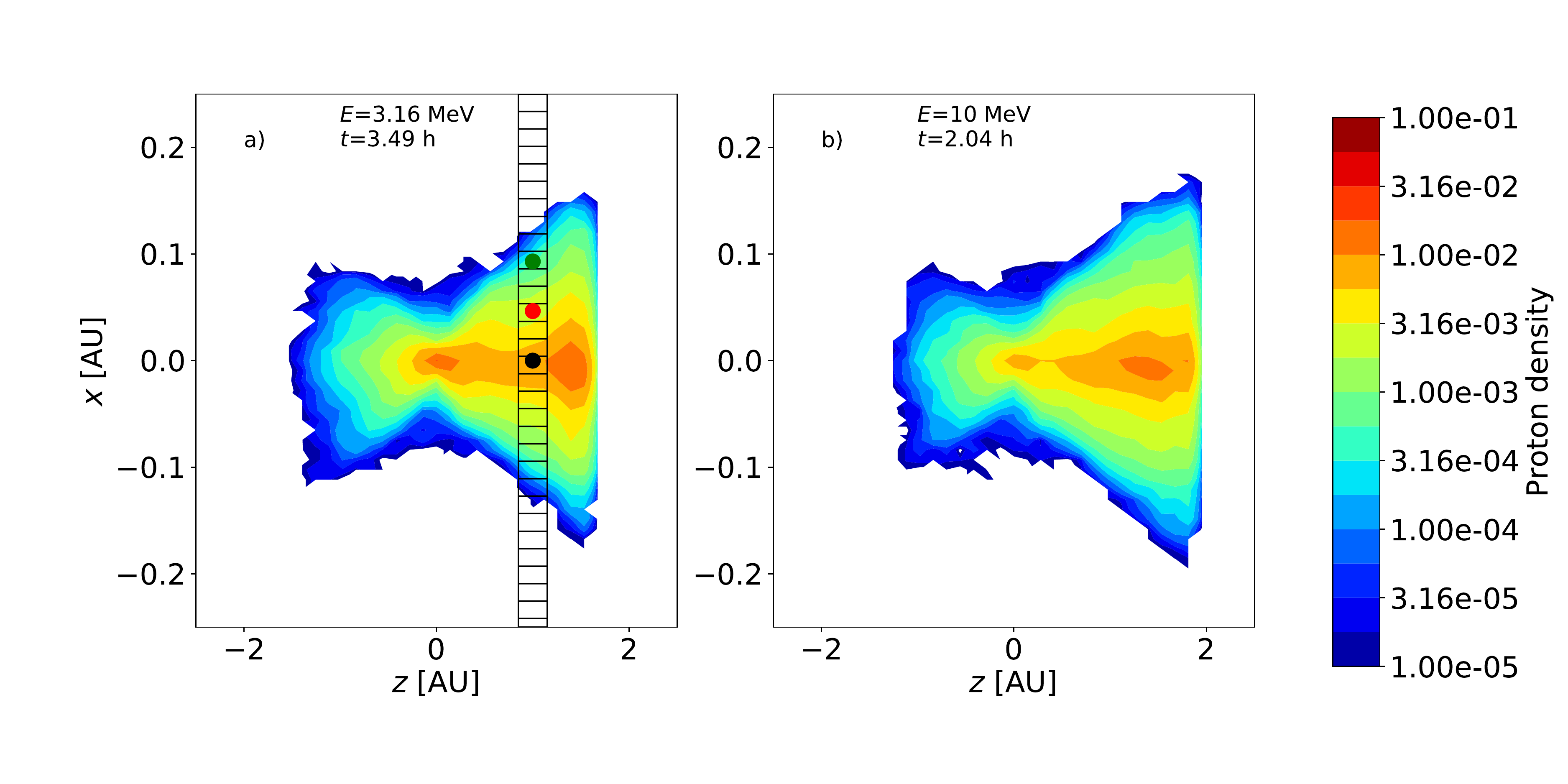}\\   
	\includegraphics[trim=5mm 20mm 5mm 5mm, clip=true,
        width=2\columnwidth]{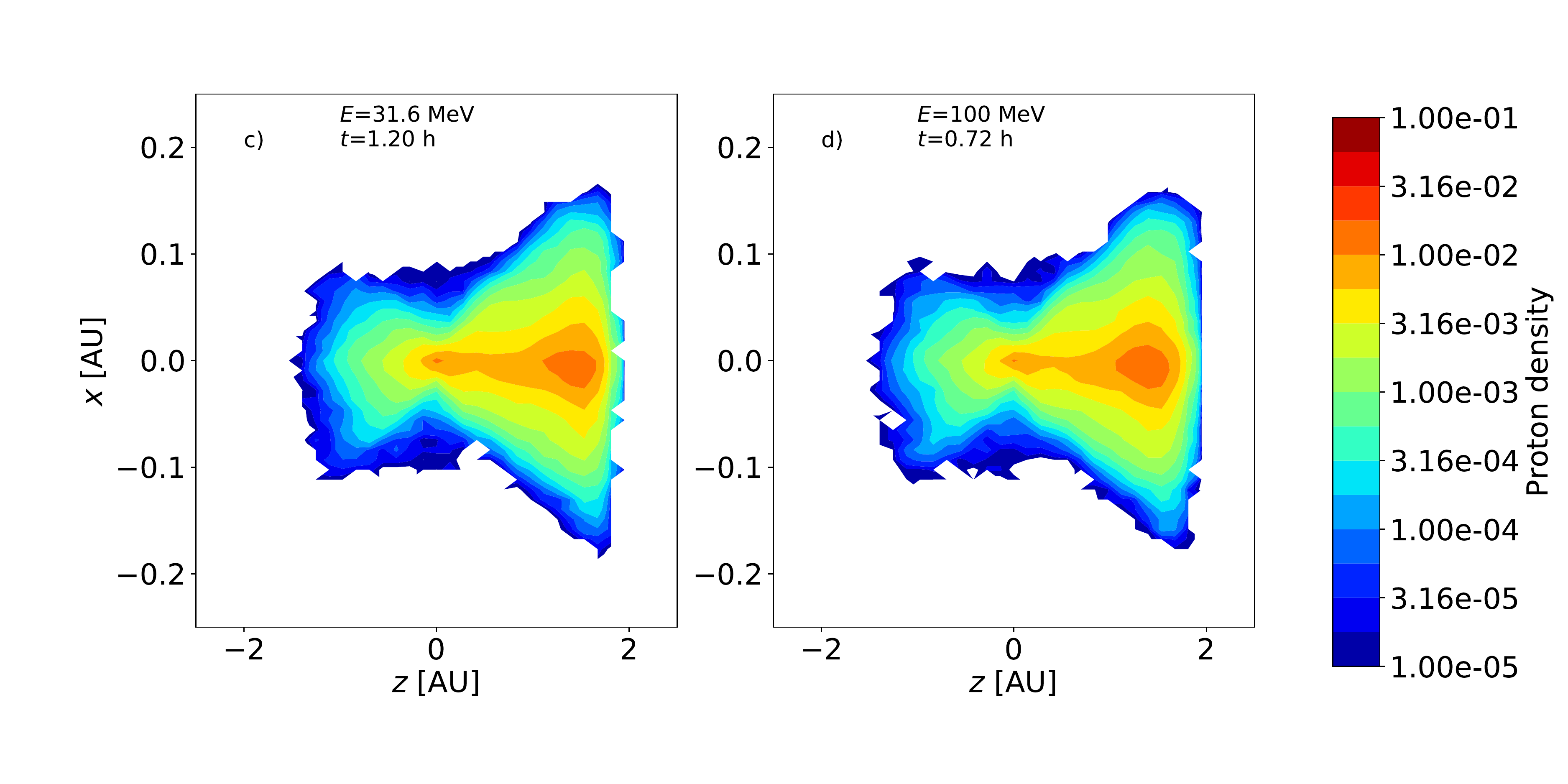} 
    \caption{Contours of the spatial distributions of protons with
      energies of 3.16, 10, 31.6 and 100~MeV, with turbulence parameters $\delta B^2/B^2=0.1$ and $q_\parallel=5/3$. The contours are given at
      the times when an unscattered proton of the given energy would
      have propagated a distance of 2 AU. The circles in panel a)
      depict positions corresponding to curves in
      Figure~\ref{fig:e_timeints}, and the hatched box at 1~AU the
      region used for determining $\sigma_\perp^2(z=1~AU, t)$.
    \label{fig:econtours}}
\end{figure*}

In this Section, we will first view the qualitative behaviour of the
particle population along and across the mean field direction for
different particle and turbulence parameters. We will then proceed to
quantify the evolution of the particle population's extent, using
simple considerations presented in
Appendix~\ref{sec:width-distr-simple}.

As a first step in forming a view on the evolution of particle
population along and across the mean field, we present a contour plot
of the proton density in Figure~\ref{fig:econtours}, for turbulence
with $\dbtbt=0.1$ and $q_\parallel=5/3$ for four different proton
energies. The distributions are integrated in $y$-direction to
  reduce numerical noise in the contours. The distributions are shown
at times when a non-scattered proton of the given energy would have
propagated the distance of $z_{ns}=2$~AU, $t=z_{ns}/v$. Thus, the four
panels depict the distribution of particles of different energies that
would have travelled the same distance, unscattered, in a constant
background magnetic field. The sharp decrease and narrowing of
  the particle distribution near $z_{ns}=2$~AU demonstrates the finite
  propagation distance of the particles and the longer distance
  travelled by particles along meandering field lines to larger
  cross-field distances in the $x$-direction. The decrease starts
  already before $z=2$~AU, as some of the first particles have
  experienced scattering due to the small-scale turbulence.

As can be seen in Figure~\ref{fig:econtours}, the particles spread
rapidly from their point of origin in the cross-field direction
(vertical axis): the front of the fastest propagation particles at
$z=2$~AU is much wider than at $z=0$~AU. A small number of
backscattered particles have advanced to the region $z<0$~AU (on
horizontal axis), and show also expansion in the cross-field direction
faster than that at $z=0$~AU.  As discussed in \citet{LaEa2013b}, the
resulting butterfly-shape of particle distribution cannot be obtained
by a simple diffusive spreading of particles. In addition, the
distribution of the particles at different distances along the field
direction is qualitatively very similar at all energies. This suggests
that the particles with different energies are propagating along the
same pathways.

\begin{figure*}
  \includegraphics[trim=5mm 20mm 5mm 5mm, clip=true,
  width=2\columnwidth]{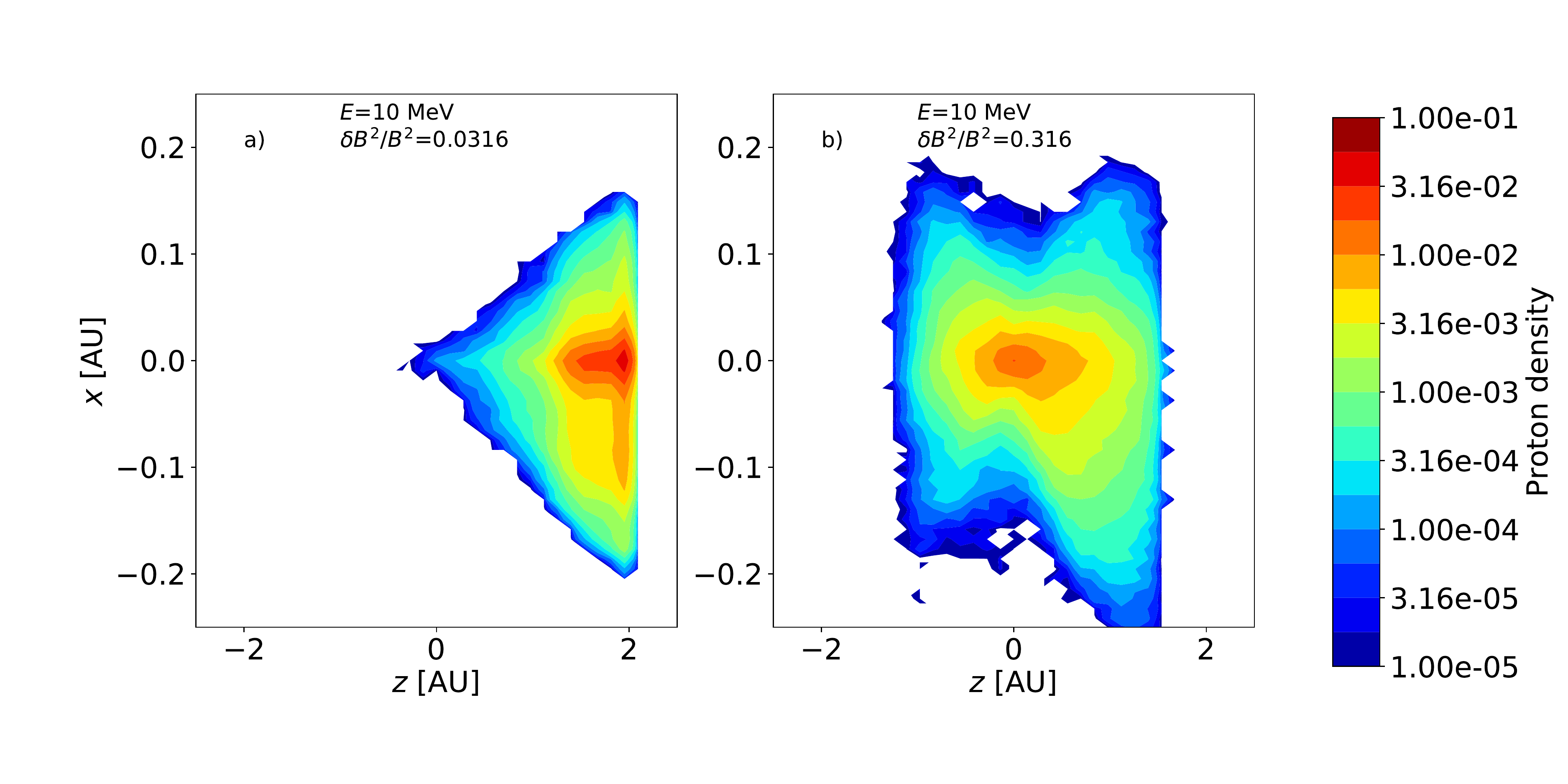}
    \caption{Contours of the spatial distributions of 10~MeV protons with
      turbulence amplitudes 0.00316 and 0.316, and $q_\parallel=5/3$. The contours are given
      at the time when the times when an unscattered proton of the
      given energy would have propagated a distance of 2 AU, at $t=2.04$~h.
      \label{fig:bcontours}}
\end{figure*}

In Figure~\ref{fig:bcontours}, we show the distribution of 10~MeV
protons in turbulence of different strength, $\dbtbt=0.0316$ and
$\dbtbt=0.316$ in panels a) and b), respectively. The distributions
are again shown at the time when an unscattered particle would have
reached 2~AU along a uniform magnetic field. The distributions differ
significantly at $z<0$~AU where the number of backscattered particles
decreases as the turbulence weakens, with almost no backscattered
particles for the weakest turbulence case, $\dbtbt=0.0316$ (Figure
\ref{fig:bcontours}~a)). On the other hand, at $z\sim 2$~AU the
particles propagating in weaker turbulence (\ref{fig:bcontours}~a))
have progressed further in parallel direction, remaining in a more
coherent pulse. These differences are due to the dependence of
parallel scattering rate on the turbulence amplitude. For 10~MeV
protons, the quasi-linear parallel mean free path, obtained from the
simulated particles, is $\lambda_\parallel=0.29$~AU for the case of
$\dbtbt=0.316$ (Figure~\ref{fig:bcontours}~b)), consistent with
considerable scattering along the mean magnetic field by the time the
particles have propagated for a time time corresponding to
scatter-free propagation of 2~AU. For the case of $\dbtbt=0.0316$, the
mean free path is $\lambda_\parallel=7.6$~AU, resulting in the almost
scatter-free propagation depicted in
Figure~\ref{fig:bcontours}~a). For the intermediate case,
$\dbtbt=0.1$, depicted in Figure~\ref{fig:econtours}~b), the mean free
path is $\lambda_\parallel=1.3$~AU, which can be seen in some parallel
diffusion and backscattering of the particles.

The cross-field extent of the particle distribution also appears to
depend on the turbulence amplitude, as seen when comparing
Figures~\ref{fig:bcontours}~a), \ref{fig:econtours}~b) and
\ref{fig:bcontours}~b), for the weak, intermediate and strong
scattering conditions, respectively, for 10~MeV protons. The
distribution is clearly narrower at lower turbulence amplitude
turbulence at $z=0$~AU. A similar trend is seen also at larger
distances along the mean field line direction: at $z=2$~AU, the weak
turbulence case, as presented in Figure~\ref{fig:bcontours}~a), the
particles are much more concentrated on a narrow cross-field extent
than in the more turbulent cases presented in
Figures~\ref{fig:econtours}~b) and \ref{fig:bcontours}~b).

\begin{figure}
  \includegraphics[
  width=\columnwidth]{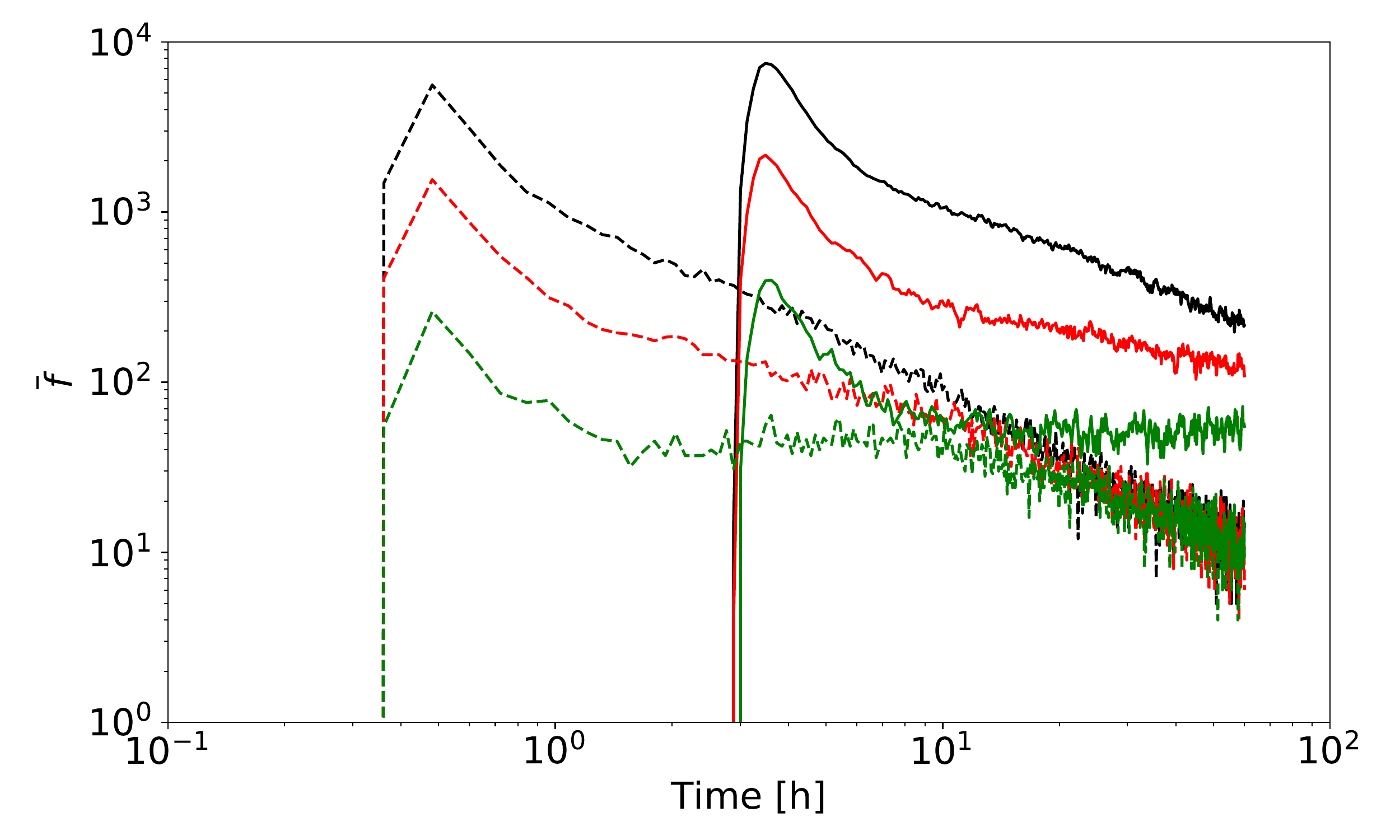}
  \caption{The time evolution of the distribution of protons at $z=1$~AU and $x=0,$~0.046 and 0.093~AU from (black, red and green curves, respectively), for 1 and 100~MeV (solid and dashed curves, respectively), and $\delta B^2/B^2=0.1$ and $q_\parallel=5/3$. \label{fig:e_timeints}}
\end{figure}

In Figure~\ref{fig:e_timeints}, we show the time development of the
particle density, as it would be measured by spacecraft situated at
the black, red and green circles in Figure~\ref{fig:econtours}~a), for
1 and 100 MeV protons (solid and dashed curves), respectively. As can
be seen, the particle density decreases strongly when the measuring
point is moved from the black circle, connected along the mean
magnetic field to the particle source, across the mean field
direction, to the red and green circles. However, the ratio of
intensities at different locations appears to be initially independent
of time (at times $<$~1~h for 100~MeV protons, and $<$~10~h for 1~MeV
protons), and only at later times the intensities at different
locations begin to converge. This behaviour of initially
time-independent cross-field distribution is seen at both energies in
Figure~\ref{fig:e_timeints}. Furthermore, the temporal behaviour in
the initial phase of the simulations, when scaled with the particle
velocity, is very similar. Thus, the initial cross-field spreading of
the particles seems to be independent of both energy and
velocity-scaled time. At later times, the intensities at different
locations begin to converge, as the width of the particle population increases.

To analyse the particle cross-field extent quantitatively, we study
the temporal evolution of the cross-field variance of the particles,
$\sigma_\perp^2(z,t)$, as a function of time at different distances
from the particle source. As discussed in the previous section, we
define the cross-field variance for particles at a given distance
along the mean field direction using Equation~(\ref{eq:fit1}), so that
for example the variance $\sigma_\perp^2(z=1\mathrm{ AU},
t=3.49\mathrm{ h})$ is determined from the particles within the
hatched box in Figure~\ref{fig:econtours}~a).

\begin{figure}
  \includegraphics[trim=0mm 32mm 15mm 35mm,clip=true,
  width=\columnwidth]{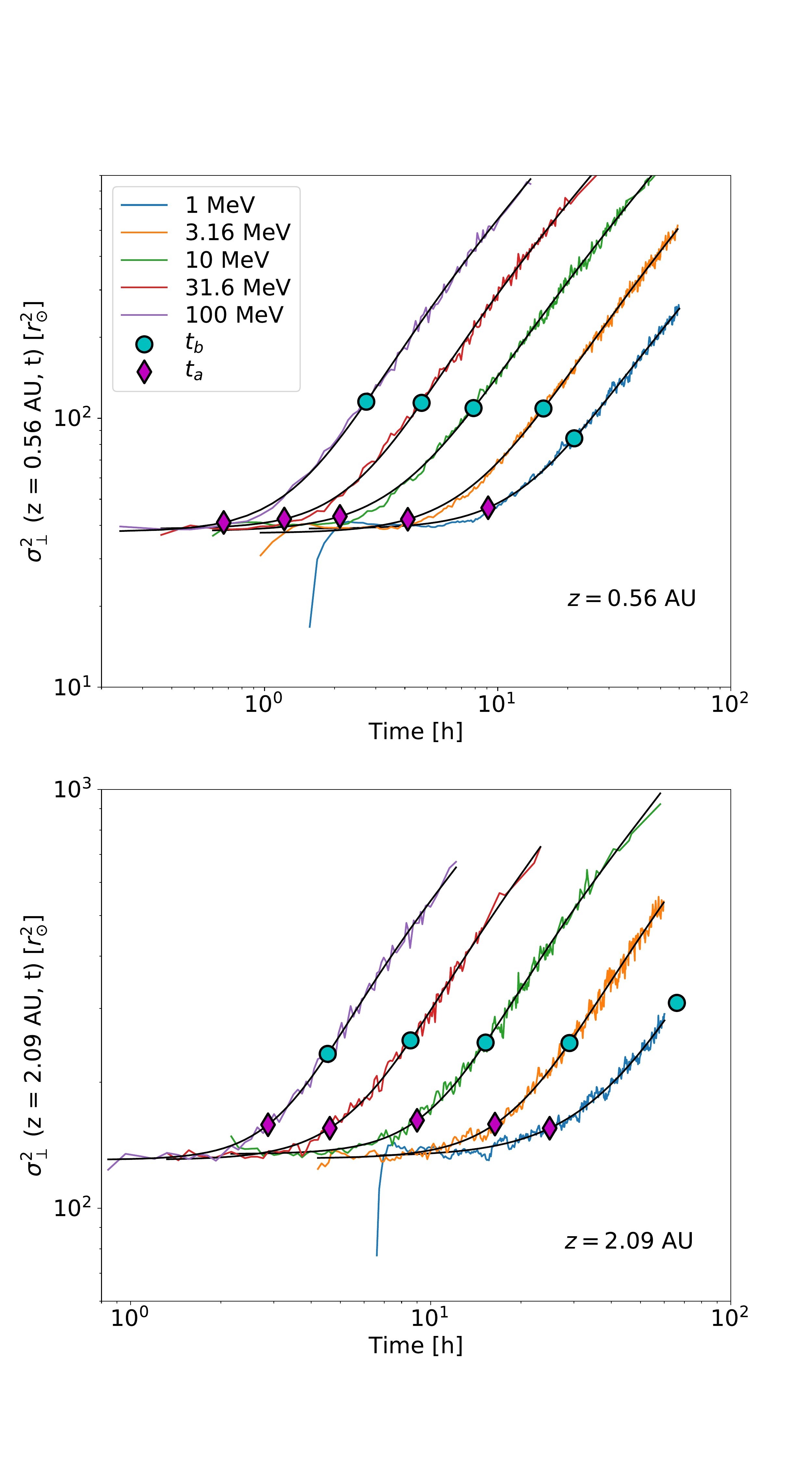}
  \caption{The mean square width of protons at 0.56~AU (top panel) and
    2.09~AU along the mean field direction from the source
    location. The black curves show fits made using
    Equation~\ref{eq:fit1}, the magenta diamond the time $t_a$ and the
    cyan circle the time $t_b$. The turbulence parameters are $\delta
    B^2/B^2=0.1$ and $q_\parallel=5/3$.\label{fig:e_timesigmas}}
\end{figure}

In Figure~\ref{fig:e_timesigmas}, we show $\sigma_\perp^2(z,t)$ at
$z=$0.56~AU (top panel) and $z=$2.09~AU (bottom panel), as a function
of time, for 1, 3.16, 10., 31.6 and 100~MeV protons, from right to
left, respectively. The black curves represent a fitted function that
is discussed below. As we can see, the temporal behaviour of the
cross-field variance of the particles can be decomposed to three time
periods. The first time period, visible only for 1~MeV protons in
Figure~\ref{fig:e_timesigmas} (blue, rightmost curve) demonstrates
fast increase of the variance to a constant level (at times $<2$~h in
the top panel). This is caused by the fact that the particles with
very small cross-field deviation are likely to have the shortest
path-lengths along the meandering field lines, and thus the first
particles will arrive earlier to the well-connected location ($x=0$~AU
in Figures~\ref{fig:econtours} and~\ref{fig:bcontours}) than to
larger cross-field deviations. The effect of the length of the
meandering fieldlines on the particle onsets at different spatial
locations will be investigated in a separate study.

The first time period is short, and due to finite time sampling of the
particle locations within the simulations, it is not visible at higher
energies. After the initial fast rise, the variance remains at a
constant level for a considerably long time, at about
$\sigma_\perp^2(z,t)=38\;\mathrm{r}_\odot^2$ at z=0.56~AU, and
$130\;\mathrm{r}_\odot^2$ at z=2.09~AU. The constant level is
independent of energy, and increases with distance along
the mean magnetic field direction (top vs bottom panel of
Figure~\ref{fig:e_timesigmas}).

Subsequently, the cross-field extent of the particle population begins
to increase and finally reaches a time-asymptotic
$\sigma_\perp^2\propto t$ trend, consistent with diffusive cross-field
spreading of the particles. Both the onset time and the time profile
of the change from constant to the diffusive phase depend on both the
particle energy and the distance along the $z$ axis. The simplest
method of obtaining the time when the cross-field diffusion becomes
significant, $t_a$, is to use time at which the time-asymptotic
straight line $\sigma_\perp^2\propto t$ intersects the initial
constant value of $\sigma_\perp^2$ level (the magenta and black dashed
lines in Figure~\ref{fig:e_fitexamples}, respectively).

The simplest approach to describe mathematically the temporal
behaviour of the cross-field variance presented in
Figure~\ref{fig:e_timesigmas} would be to consider a diffusive
spreading of particles from the meandering field lines with constant
diffusion coefficient, which would result in variance given by
Equation~(\ref{eq:simplediff}). While such a model is easy to
implement and can be used as the first approach, as in, e.g.,
\citet{LaEa2016parkermeand}, we will consider here an improvement to
the modelling of the transition, based on the work by
\cite{LaDa2017decouple}. In Appendix~\ref{sec:width-distr-simple}, we
derive a model with time-dependent diffusion coefficient, which will
justify fitting the variance with functional forms such as
Equation~(\ref{eq:timedepdiff}). In this work, we will use
\begin{equation}
  \label{eq:fit1}
  \sigma_\perp^2(z,t)=\sigma_{\perp 0}^2(z)+\frac{\sigma_{\perp 1}^2(z)\,t/t_0}{1+\left[t_b(z)/t
\right]^{\alpha(z)-1}}.
\end{equation}
where $\sigma_{\perp 0}^2(z)$ is the early-time constant cross-field
variance of the particles at distance $z$, $t_0=1$~h the unit of time,
$t_b(z)$ is the onset time of the time-asymptotic diffusive regime,
and $\alpha(z)$ the power law index that describes the fast
spreading of the particle population before $t_b$. The time-asymptotic
diffusive rate of change of the cross-field variance of the particle
population is given by $\sigma_{\perp 1}^2(z)/t_0\equiv
2\,\kappa_\perp(t\gg t_b)$, with $\kappa_\perp(t)$ the time-dependent
particle diffusion coefficient (see
Appendix~\ref{sec:width-distr-simple}). Using the parameters in
Equation~(\ref{eq:fit1}), $t_a=\sigma_{\perp 0}^2/\sigma_{\perp 1}^2\;
t_0$. The times $t_a$ and $t_b$, and their relation to the
$\sigma_\perp^2(z,t)$ curve and its asymptotes are shown in
Figure~\ref{fig:e_fitexamples}, and are further discussed in
Appendix~\ref{sec:width-distr-simple}.

It should be noted that Equation~(\ref{eq:fit1}) is of the same form
as the fitting function used in \citet{LaDa2017decouple}, at the limit
of $t\gg t_1$ in their formulation. Here, we exclude the early
behaviour, $t\lesssim t_1$ discussed in \citet{LaDa2017decouple}, as
displacements of order Larmor radius are too small to be seen in our
current analysis.

We fit Equation~(\ref{eq:fit1}) to the cross-field variances from our
simulations at different distances $z=$0.3--3~AU, and study the
dependence of the fit parameters on $z$, the particle energy and the
turbulence properties. We exclude fits where the relative error of any
of the fit parameters in Equation~(\ref{eq:fit1}) exceeds 10\%. As an
example, we show the fits of the simulation results with
Equation~(\ref{eq:fit1}) in Figure~\ref{fig:e_timesigmas} with black
curves, with $t_b$ shown with a cyan circle. As can be seen, Equation
(\ref{eq:fit1}) mostly fits the simulation results well at all
energies on both locations.

\begin{figure*}
  \includegraphics[
  width=2\columnwidth]{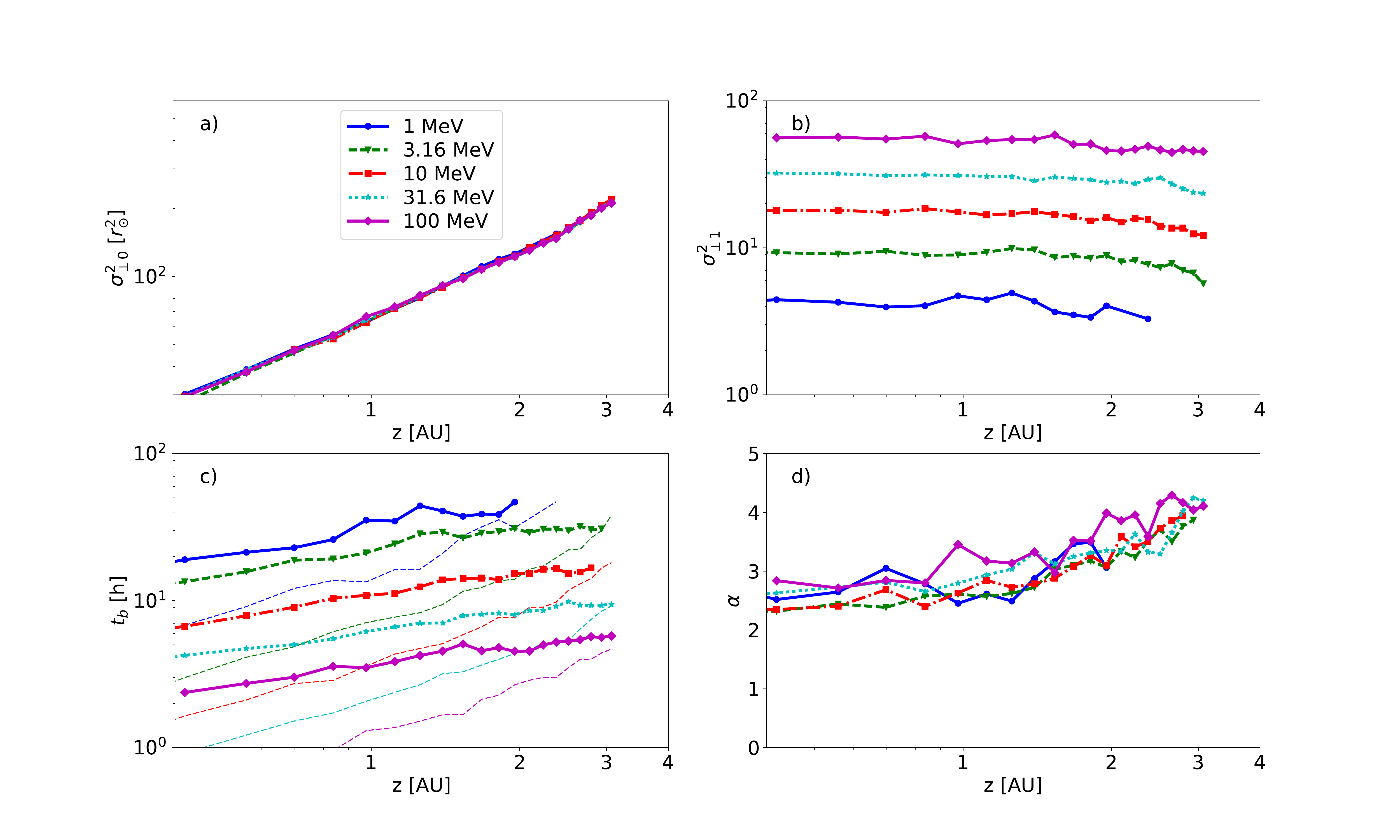}
  \caption{The fitting parameters of the Equation~(\ref{eq:fit1}) to
    the cross-field particle distribution variances at different
    distances, for protons of energies 1--100~MeV propagating in
    turbulence with $\delta B^2/B^2=0.1$ and
    $q_\parallel=5/3$. a) The early-time cross-field variance of the distribution,
    $\sigma_{\perp 0}^2$; b) the asymptotic long-time behaviour
    $\sigma_{\perp 1}^2$; c) The turnover time $t_b$ (thick curves)
    and the asymptotic behaviour changing time $t_a$ (thin dashed curves);
    d) The power law index $\alpha$.
    \label{fig:efitsq167r2}}
\end{figure*}

In Figure~\ref{fig:efitsq167r2}, we show the evolution of the fit
parameters as a function of $z$ for five proton energies for
turbulence parameters $\delta B^2/B^2=0.1$ and
$q_\parallel=5/3$. Panel a) shows the early-time constant cross-field
variance $\sigma_{\perp 0}^2$, which increases approximately linearly
as a function of distance. The $\sigma_{\perp 0}^2(z)$ curves for
different energies overlay each other almost perfectly, suggesting
that the spread of particles is caused by the diffusive spreading of
the meandering magnetic field lines, rather than particle scattering
in space across the mean field.

In Figure~\ref{fig:efitsq167r2} b), we show $\sigma_{\perp 1}^2$,
which describes the time-asymptotic diffusive spreading rate of the
particles. As can be seen, $\sigma_{\perp 1}^2$ is clearly
energy-dependent, and almost independent of $z$, as expected for
diffusive spreading of particles. We determined that $\sigma_{\perp
  1}^2$ is roughly proportional to the particle velocity, and is also
consistent with the theoretical dependence $\kappa_\perp\propto
R^{10/9}$ of the particle diffusion coefficient on rigidity
\citep{Matthaeus2003,Shalchi2004analytic}.

In Figure~\ref{fig:efitsq167r2} c), we show $t_b$, which represents
the onset time of the diffusive spreading of particles. A closer
inspection shows that the onset time is roughly proportional to the
inverse of velocity. Similar scaling was found by
\citet{LaDa2017decouple}, who noted that the transition timescale from
the slow to fast diffusive spreading at $z=0$~AU is proportional to
the parallel scattering timescale $\tau_\parallel\propto v^{-2/3}$.

However, as can be seen in Figure~\ref{fig:efitsq167r2}, the onset
time $t_b$ is not independent of the distance $z$, but increases
slowly, inconsistent with the assumptions of the simple diffusion
model described in Equation~(\ref{eq:timedepdiff}). The inconsistency
can be caused by the finite propagation time of the particles to
distance $z$, not taken into account in
Equation~(\ref{eq:timedepdiff}).

In Figure~\ref{fig:efitsq167r2} c), we also show the asymptotic time
$t_a$, with the thin dashed curves. As can be seen, $t_a$ approaches and
surpasses $t_b$ at larger distances. As discussed in
Appendix~\ref{sec:width-distr-simple}, the determination of $t_b$ and
$\alpha$ are unreliable when $t_a$ approaches $t_b$, and the unreliable
values of $t_b$ are not shown in Figure~\ref{fig:efitsq167r2}~c).

Figure~\ref{fig:efitsq167r2} d) shows the parameter $\alpha$,
which describes the fast spreading of the particles
until the time-asymptotic diffusive phase is reached. As interpreted
by \citet{LaDa2017decouple}, the fast spreading begins when
the particles cease being well tied to their field lines. The onset
time of the fast spreading phase, denoted by $t_1$ in
\citet{LaDa2017decouple}, however, is masked by the early-time
$\sigma_{\perp 0}^2$ (see Figure~\ref{fig:e_timesigmas}), and can only be
quantified using the method described in \citet{LaDa2017decouple}. The
$\alpha$ in Figure~\ref{fig:efitsq167r2}~d) shows no clear
energy dependence, consistent with
\citet{LaDa2017decouple}. 

\begin{figure*}
  \includegraphics[
  width=2\columnwidth]{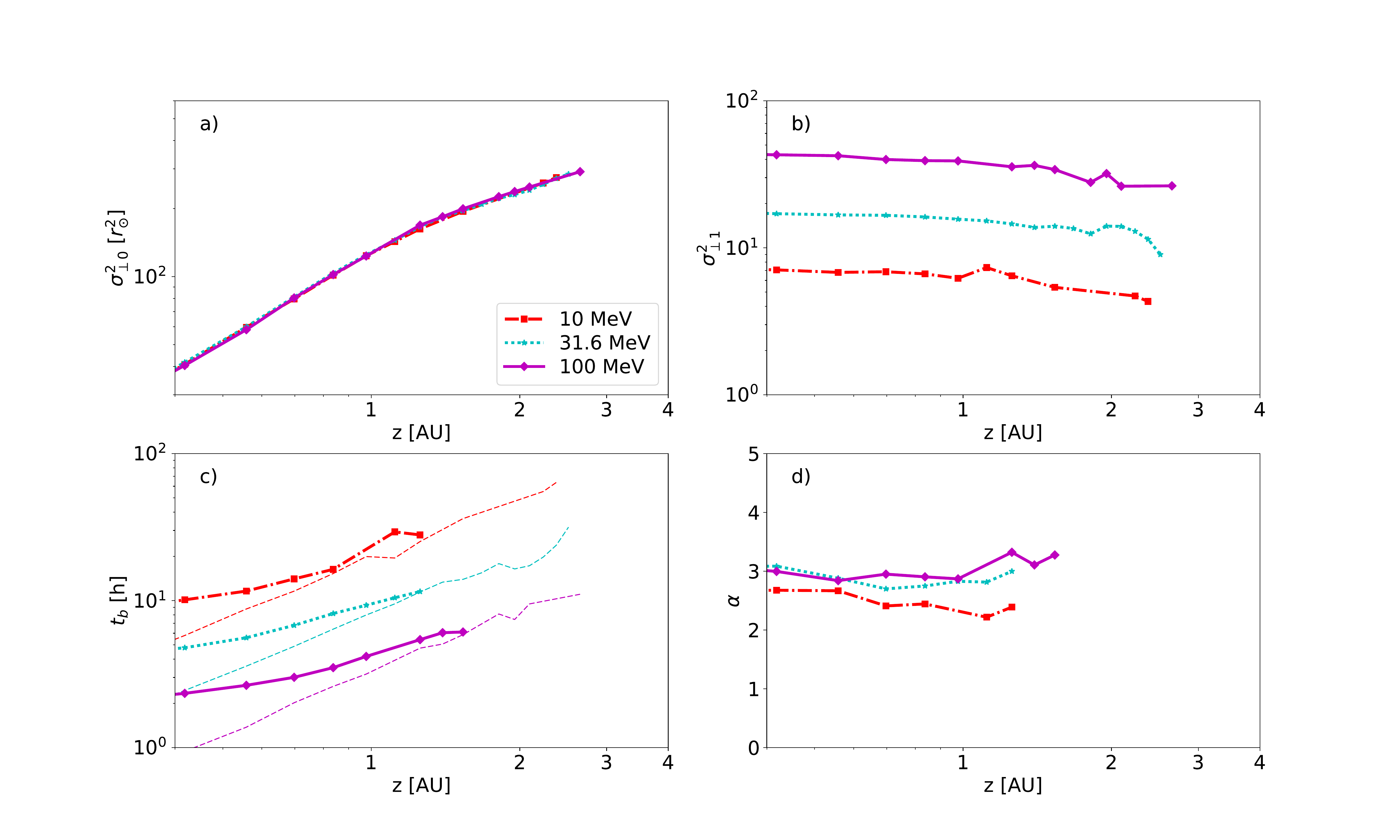}
  \caption{The fitting parameters of the Equation~(\ref{eq:fit1}) to
    the cross-field particle distribution variances at different distances, for proton
    energies 10--100~MeV propagating in turbulence with $\delta B^2/B^2=0.1$ and
    $q_\parallel=1$. a) The early-time cross-field variance  of the distribution,
    $\sigma_{\perp 0}^2$; b) the asymptotic long-time behaviour
    $\sigma_{\perp 1}^2$; c) The turnover time $t_b$ (thick curves)
    and the asymptotic behaviour changing time $t_a$ (thin dashed curves); d) The
    power law index $\alpha$.
    \label{fig:efitsq100r1}}
\end{figure*}

In Figure~\ref{fig:efitsq100r1}, we show the simulation fit parameters
for particles in turbulence with slab spectral index
$q_\parallel=1$. 
%
%
The fit of the variance with Equation~(\ref{eq:fit1}) could be
successfully performed only for energies in the range
E=10-100~MeV. For lower energy protons, the cross-field variance
remained almost constant for the simulation period of 60~hours.

In panel a) we can see that the early-time, constant cross-field variance
$\sigma_{\perp 0}$ is independent of energy also for
$q_\parallel=1$. This energy-independence  was evident also for energies
$<10$~MeV, for which the fitting of Equation~(\ref{eq:fit1}) was not
successful. In panel b), we note that $\sigma_{\perp 1}^2$ is again
almost independent of $z$. However, its dependence on energy is
different than in the case $q_\parallel=5/3$
(Figure~\ref{fig:efitsq167r2}~b)): the dependence is now close to
$\sigma_{\perp 1}^2\propto p^{1.5}$. This dependence is again close to
the NLGC prediction $\lambda_\perp\propto \lambda_\parallel^{1/3}$ which,
together with the quasilinear theory result $\lambda_\parallel \propto
R^{2-q_\parallel}$ and $q_\parallel=1$ gives
$\kappa_\perp=v\lambda_\perp/3\propto p^{4/3}$ at non-relativistic limit
\citep{Matthaeus2003,Shalchi2004analytic}.

In panel c) of Figure~\ref{fig:efitsq100r1}, we show the behaviour of
the onset time $t_b$ as a function of energy and distance. Compared to
the $q_\parallel=5/3$ case, the dependence of the onset time on energy
is stronger, with $t_b\propto 1/v^{3/2}$. This is in contradiction
with the suggestion that the timescale would scale as the parallel
scattering time \citep{LaDa2017decouple}, as this timescale,
$\lambda_\parallel/v$, would be constant for $q_\parallel=1$. The
discrepancy may be due to the strong parallel scattering of the
particles for the $q_\parallel=1$ case. In a strongly diffusive
environment, the particles propagate in timescales $t\propto z^2$,
thus much slower for the scatter-dominated case of $q_\parallel=1$, as
compared to the almost scatter-free case with $q_\parallel=5/3$. The
slower parallel propagation implies that the finite propagation time
effects, not accounted for in the simple diffusion model, are much
more pronounced for the $q_\parallel=1$ case.

\begin{figure*}
  \includegraphics[
  width=2\columnwidth]{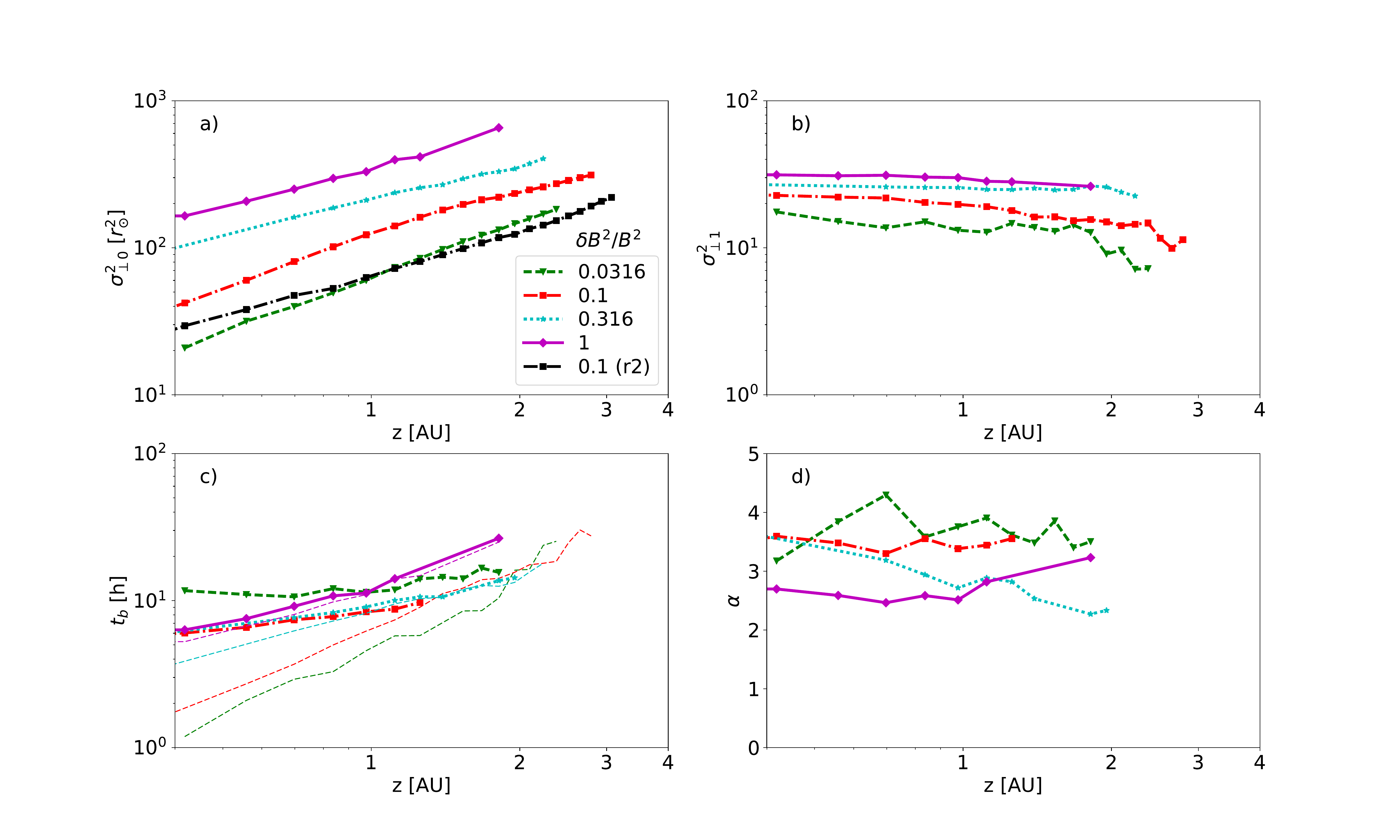}
  \caption{The fitting parameters of the Equation~(\ref{eq:fit1}) to
    the 10 MeV proton cross-field variances at different distances
    propagating in turbulence with different amplitudes amplitudes
    $\dbtbt$ and $q_\parallel=5/3$. a)
    The early-time cross-field variance of the distribution, $\sigma_{\perp 0}^2$, with
    the black dash-dotted curve showing a different turbulence
    realisation as compared to the red dash-dotted curve, for
    $\dbtbt=0.1$; b) the asymptotic long-time behaviour $\sigma_{\perp
      1}^2$; c) The turnover time $t_b$ (thick curves) and the
    asymptotic behaviour changing time $t_a$ (thin dashed curves); d)
    The power law index $\alpha$.
    \label{fig:bfitsq167r1}}
\end{figure*}

In Figure~\ref{fig:bfitsq167r1}, we investigate the effect of varying
turbulence amplitudes on the 10~MeV proton distribution extent across
the mean field, with $q_\parallel=5/3$, in the same format as in
Figures~\ref{fig:efitsq167r2} and~\ref{fig:efitsq100r1}. The
turbulence at different amplitudes is obtained so that the random
phases and polarisations of the Fourier modes were the same
for each simulation, and only the amplitudes of the Fourier modes were
changed.

Figure~\ref{fig:bfitsq167r1} a) shows the dependence of the early-time
constant cross-field extent $\sigma_{\perp 0}^2$ on the turbulence
amplitude, from $\dbtbt=0.0316$ (dashed green curve) to $\dbtbt=1$
(solid magenta curve). The initial variance is proportional to the
distance $z$ along the mean field direction at all turbulence
amplitudes, again consistent with particles propagating on diffusively
meandering field lines. We found that $\sigma_{\perp 0}^2/z\propto
\sqrt{\dbtbt}$, which is consistent with the dependence of the field
line diffusion coefficient on the amplitude of the 2D-turbulence
\citep{Matthaeus1995}.

In Figure~\ref{fig:bfitsq167r1}~a) we also show the early-time
cross-field variance $\sigma_{\perp 0}^2$ for a simulation with
$\dbtbt=0.1$ with random phases and polarisations different from the
ones used in the other simulations presented in
Figure~\ref{fig:bfitsq167r1}, with black dash-dotted curve. As can be
seen, the two realisations with $\dbtbt=0.1$ (black and red
dash-dotted curves) differ by factor~2. Thus, different realisations
can locally produce significantly different turbulence conditions,
which can affect parametric comparison of particle and field line
dynamics in simulated turbulence. For this reason, it is important to
keep consistency in the simulated fields when comparing, e.g., the
effect of turbulence ampitudes only on the particle transport, as in
Figure~\ref{fig:bfitsq167r1}.

In Figure~\ref{fig:bfitsq167r1} b) we see that the time-asymptotic
diffusive spreading rate does not depend strongly on the turbulence
amplitude. Upon closer inspection, we find that $\sigma_{\perp 1}^2$
from our fits is proportional to $(\dbtbt)^{0.25}$, which is slightly
flatter than the dependence given by the NLGC value
$\kappa_\perp\propto (\dbtbt)^{2/3} \kappa_\parallel^{1/3}$
\citep{Shalchi2004analytic}, which for quasilinear theory
$\kappa_\parallel\propto (\dbtbt)^{-1}$ \citep{Jokipii1966} would give
$\kappa_\perp\propto (\dbtbt)^{1/3}$. In our simulations, the
dependence of $\kappa_\parallel$ on turbulence amplitude, however,
differed slightly from the quasilinear theory value: using our
$\kappa_\parallel$ from the simulations in the NLGC expression we get
$\kappa_\perp\propto(\dbtbt)^{0.2}$, close to the $\sigma_{\perp
  1}^2\propto(\dbtbt)^{0.25}$ obtained from our fits in Figure~\ref{fig:bfitsq167r1}~b). In addition, the
recent work by \citet{RuffoloEa2012} found a non-power law, flattening
behaviour for $\kappa_\perp$ at higher turbulence amplitudes, which
may be visible also in our simulations.

In Figure~\ref{fig:bfitsq167r1} c), we see that $t_b$ decreases
with increasing turbulence amplitude for the two lowest turbulence
amplitudes (the green dashed curve and red dash-dotted curve),
again consistent with the \citet{LaDa2017decouple} result.
Subsequently, at higher amplitudes, the onset time begins to
increase. The increase can in part be due to slow propagation times of
the more diffusive particles in higher-amplitude turbulence, as well
as due to the asymptotic time $t_a$ (the thin dashed curves)
approaching $t_b$ which, as discussed in
Appendix~\ref{sec:width-distr-simple}, renders the fitting of the
variance to Equation~(\ref{eq:fit1}) insensitive to the parameters
$t_b$ and $\alpha$. The resulting spurious behaviour can also be seen
in the fitted value of the power law index $\alpha$ in
Figure~\ref{fig:bfitsq167r1}~d).

\section{Discussion}\label{sec:discussion}

Our analysis shows that the initial cross-field propagation of
particles in turbulent magnetic fields is non-diffusive over a wide
range of turbulence and particle parameters, reinforcing the
conclusions made in the case study by \citet{LaEa2013b}. Initially
particles remain on their field lines, as demonstrated by the
energy-independent initial spreading of the particles in different
turbulence environments (Figures~\ref{fig:efitsq167r2}~a) and
\ref{fig:efitsq100r1}~a)), which is followed by a time-asymptotic,
energy-dependent diffusive spreading of the particles across the mean
field direction. The early-time constant cross-field variance phase is
seen as a very fast access of particles to wide cross-field ranges,
and it lasts for hours to tens of hours depending on particle and
turbulence parameters.

We found that the early-time cross-field extent of the particle
population, as given by the cross-field variance $\sigma_{\perp 0}^2$
in Equation~(\ref{eq:fit1}), is proportional to the distance $z$ from
the particle source along the mean field direction, and
$\sqrt{\dbtbt}$. Thus, the initial propagation phase is consistent
with particles propagating along diffusively meandering field lines in
turbulence dominated by 2D wave modes \citep{Matthaeus1995}. At the
time-asymptotic limit, the rate of cross-field spreading obtained in
our simulations can be described as diffusion with cross-field
diffusion coefficient $\kappa_\perp$ proportional to velocity (for the
$q_\parallel=5/3$ case) and $(\dbtbt)^{0.25}$, consistent with the our
current theoretical understanding of asymptotic particle cross-field
diffusion
\citep[e.g.][]{Matthaeus2003,Shalchi2004analytic,RuffoloEa2012}.
Thus, the early-time and the time-asymptotic phases of the particle
cross-field propagation in our simulations are consistent with our
theoretical understanding of the field line and particle behaviour at
the corresponding limits.

Our study shows that the transition between the early spreading along
the meandering field lines, and the time asymptotic diffusion phase
does not follow the simple diffusion picture presented in
Equation~(\ref{eq:simplediff}) in most cases studied in this work, but
suggests a delayed onset of the diffusion phase after \text{fast}
expansion, as indicated by the recent results by
\citet{LaDa2017decouple}. As discussed in
Appendix~\ref{sec:width-distr-simple}, assuming particles to diffuse
from the meandering field lines at a constant rate (magenta curve in
Figure~\ref{fig:e_fitexamples}) can result in overestimation of the
cross-field variance of the particle population by up to a factor of 2
during the transition phase between the initial and time-asymptotic
propagation phases. We find that for low-scattering conditions, the
onset of the time-asymptotic phase, $t_b$, follows approximately the
dependence on the parallel scattering rate obtained by
\citep{LaDa2017decouple}. For stronger tubulence, however, the scaling
does not hold any more, possibly due to finite propagation time
effects, which are not contained in our simple model behind the
formulation of Equation~(\ref{eq:fit1}). The particles may have also
been already decoupled from their fieldlines before reaching
distance~$z$, which would result in inaccuracy in the determination of
$t_b$, as discussed in the Appendix~\ref{sec:width-distr-simple}.

\section{Conclusions}\label{sec:conclusions}

In this work, we have studied how the turbulence and particle
properties affect the particle spreading across the mean field
direction early after the particle release. Our results show that

\begin{itemize}
\item Initially, the particles spread systematically and rapidly to a
  wide cross-field range along meandering field lines. They remain on
  the fieldlines from hours up to tens of hours for protons of
  1--100 MeV, in turbulence consistent with proton parallel mean free
  paths $\lambda_\parallel \sim 0.1-1$~AU, range consistent with those
  measured during solar energetic particle events
  \citep[e.g.][]{Palmer1982,Torsti2004} and interplanetary turbulence
  conditions
  \citep[r.g.][]{Pei2011difftens,LaEa2016parkermeand,Strauss2017perpel}.
\item The late-time behaviour of the particles is consistent with
  energy-dependent cross-field diffusion, and is consistent
  with the current theoretical understanding of the time-asymptotic
  cross-field diffusion of particles in turbulent magnetic fields
  \citep[e.g.][]{Matthaeus2003,Shalchi2004analytic,RuffoloEa2012}.
\item The transition between the initial and time-asymptotic behaviour
  can be roughly modelled as a simple spatial diffusion of particles
  from their meandering field lines. More precise modelling reveals a
  delayed decoupling of particles from the meandering field lines, as
  demonstrated in \citet{LaDa2017decouple}.
\end{itemize}

Thus, our study shows that over a wide range of turbulence and
particle parameters SEP cross-field propagation cannot be modelled by
cross-field diffusion alone early in the SEP event, but the systematic
propagation along meandering field lines must be taken into
account. Simple models, such as the one applied in
\citet{LaEa2016parkermeand}, which model the particle distribution
evolution as cross-field diffusion of particles from meandering field
lines provide a significant improvement as compared to the earlier
models diffusing particles with respect to the mean magnetic
field. Future fully consistent cross-field propagation models should
also include the timescales related to the decoupling of the particles
from the meandering field lines \citep{LaDa2017decouple}. For
early-time evolution of the particle populations, a time-dependent
diffusion coefficient, such as used in
Equation~(\ref{eq:timedepdiff}), may prove useful. We will investigate
advanced modelling of cross-field particle propagation in a future
work.

\section*{Acknowledgements}
TL and SD acknowledge support from the UK Science and Technology
Facilities Council (STFC) (grants ST/J001341/1 and ST/M00760X/1), and
the International Space Science Institute as part of international
team 297. D. Marriot was funded by the Royal Astronomical Society. This
work used the DiRAC Complexity system, operated by the University of
Leicester IT Services, which forms part of the STFC DiRAC HPC Facility
(www.dirac.ac.uk). This equipment is funded by BIS National
E-Infrastructure capital grant ST/K000373/1 and STFC DiRAC Operations
grant ST/K0003259/1. DiRAC is part of the National
E-Infrastructure. Access to the University of Central Lancashire's High
Performance Computing Facility is gratefully acknowledged.




\appendix

\section{The width of a distribution in a simple diffusion picture}\label{sec:width-distr-simple}

If we exclude the finite propagation time effects on particle
propagation, the diffusive spreading of a particle population along
and across the mean field can be described as spatial diffusion as
\begin{equation}
  \label{eq:threeddiffeq}
  \pder{n_{3D}}{t}=\nabla\cdot \hat\kappa \nabla n_{3D}
\end{equation}
where $n_{3D}$ is the density of the particles, and the non-zero
elements of the diffusion tensor $\hat\kappa$ are
$\kappa_{xx}=\kappa_{yy}\equiv\kappa_\perp$ and
$\kappa_{zz}\equiv\kappa_\parallel$. For impulsive point-source
injection, the solution to the diffusion equation is
\begin{equation}
  n_{\mathrm{3D}}(r,t)= \frac{I_0}{\left(4 \kappa_\perp^2\kappa_\parallel t\right)^{3/2}}
  \exp\left\{-\frac{ x^2+y^2}{4 \kappa_\perp t}-\frac{ z^2}{4 \kappa_\parallel t}\right\}.
  \label{eq:diffusion}
\end{equation}
Taking the second moment of this with respect to $x$, we find for the
variance in the x direction 
\begin{equation}
 \sigma_x^2(t,y,z)=2 \kappa_\perp t 
\label{eq:simplesigma}
\end{equation}
Thus, for an impulsive injection and pure diffusion, the variance of
the particle population at all $z$ grows linearly with time.

\begin{figure}
  \includegraphics[
  width=\columnwidth]{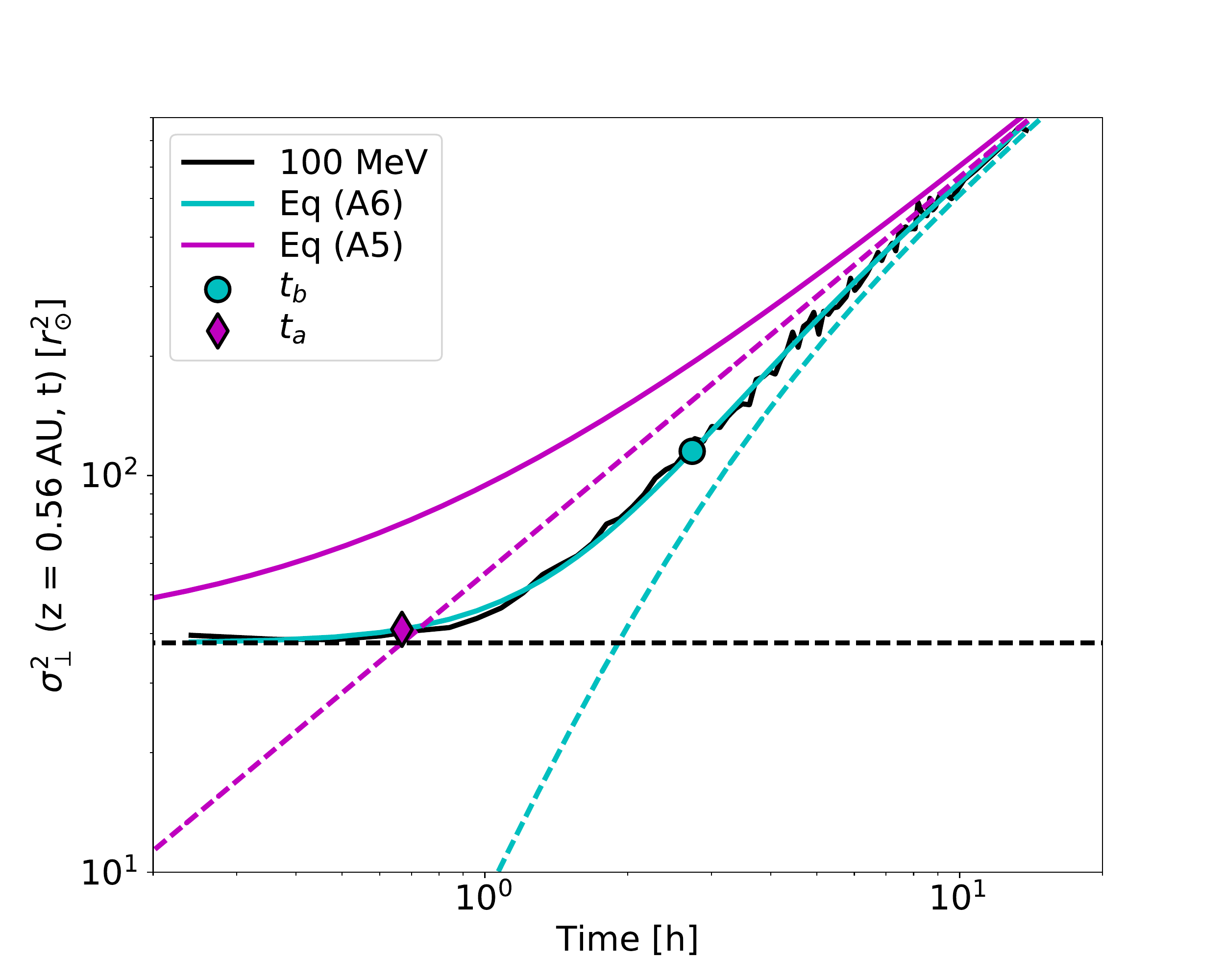}
  \caption{The cross-field variance of 100~MeV protons at
    0.56~AU, with turbulence parameters $\delta B^2/B^2=0.1$ and
    $q_\parallel=5/3$. The solid magenta curve shows the fit to
    Equation~(\ref{eq:simplediff}) and the solid cyan curve to
    Equation~(\ref{eq:timedepdiff}), using the functional shape given
    in Equation~\ref{eq:fit1}. The black dashed line shows the
    asymptotic $ \sigma_\perp^2(t,z)=\sigma_{\perp 0}$, the dashed
    cyan curve the second term in Equation~(\ref{eq:fit1}), and the
    dashed magenta curve the asymptotic $
    \sigma_\perp^2(t,z)=\sigma_{\perp 1} t/t_0$. The magenta diamond
    shows the intercept of the asymptotes, $t_a=\sigma_{\perp
      0}/\sigma_{\perp 1} t_0$, and the cyan circle shows the
    $t_{b}$.  \label{fig:e_fitexamples}}
\end{figure}

\citet{LaEa2013b} showed that cross-field diffusion does not describe
the particle cross-field distribution early in the event and concluded
that the particles follow the meandering field lines systematically,
and diffuse from them slowly. To describe such behaviour, we consider
as the first approach a simple model where the particles are initially
distributed in the cross-field $x$-direction on a Gaussian
distribution with $\sigma^2=2\,z\,D_{FL}$, where $D_{FL}$ is the
diffusion coefficient for the field lines, mimicking the cross-field
distribution of particles of particles that propagate along
turbulently meandering field lines. Subsequently, the particles spread
diffusively in the $x$-direction from their field lines. As we ignore
the propagation along the field lines in this simple model, we can
describe the evolution of the particle density as
\begin{equation}
  \label{eq:oneddiffeq}
  \pder{n_{1D}(x,z,t)}{t}=\pder{}{x}\kappa_\perp \pder{n_{1D}(x,z,t)}{x}
\end{equation}
which, for the Gaussian initial condition yields for the cross-field
variance 
\begin{equation}\label{eq:simplediff}
  \sigma_\perp^2(t,z)=2\,z\,D_{FL}+2\kappa_\perp t.
\end{equation}
We show the behaviour of the cross-field variance given by
Equation~(\ref{eq:simplediff}) in Figure~\ref{fig:e_fitexamples} with
the solid magenta curve, comparing it with a simulation case with 100~MeV
protons at 0.56~AU along the mean field direction, with turbulence
variance $\dbtbt=0.1$ and $q_\parallel=5/3$. The first and second term
in Equation~(\ref{eq:simplediff}) are shown with the black and magenta
dashed curves, respectively, and the intercept time of the terms,
$t_a$, shown with the magenta diamond. As can be seen, the magenta
curve does not trace the simulation result, solid black curve, well
during the transition phase.

Recently, \citet{LaDa2017decouple} noted that the decoupling of
particles from their initial field lines is initially slow, and only
at later times rapidly converges to the time-asymptotic diffusion
trend. Such a behaviour can be mimicked by allowing time-dependence
for the particle diffusion coefficient, with
$\kappa_\perp(t)=\kappa_{\perp0} \,T(t)$. Substituting
$\mathrm{d}\tau=T(t)\mathrm{d}t$ we can solve the
Equation~(\ref{eq:oneddiffeq}) for $\kappa_\perp(t)$, which results,
for impulsive point source, in variance
$\sigma_{x}^2(t)=2\kappa_{\perp 0} \tau(t)$, and for a Gaussian
initial condition in
\begin{equation}\label{eq:timedepdiff}
  \sigma_\perp^2(t,z)=2\,z\,D_{FL}+2\kappa_{\perp 0} \tau(t).
\end{equation}
We demonstrate the use of the solution given by
Equation~(\ref{eq:timedepdiff}), parametrised as in
Equation(\ref{eq:fit1}), in Figure~\ref{fig:e_fitexamples}, for the
simulated case of 100~MeV protons at 0.56~AU along the mean field
direction. turbulence variance $\dbtbt=0.1$ and $q_\parallel=5/3$. As
can be seen, the Equation~(\ref{eq:timedepdiff}) describes the time
evolution of the simulation results well.

When fitting the $\sigma_\perp^2(z,t)$ from the simulations to
Equation~(\ref{eq:fit1}), the difference between the simple approach,
Equation~(\ref{eq:simplesigma}), and Equation~(\ref{eq:fit1}), is not
always as clear as in Figure~\ref{fig:e_fitexamples}. If the $t_b\ll
t_a$, the second term in Equation~(\ref{eq:fit1}) approaches
$\sigma_{\perp 1}^2(z)\,t/t_0$, and is longer sensitive to parameters
$t_b$ and $\alpha$. The $t_b\ll t_a$ implies that the onset of the
diffusive phase has taken place before the particles have reached the
distance $z$.

To avoid showing the spurious fit parameters, we show the fit
parameters $t_b$ and $\alpha$ only when $t_b\ge t_a$ in
Figures~\ref{fig:efitsq167r2},
\ref{fig:efitsq100r1}~and~\ref{fig:bfitsq167r1}.


\bsp	
\label{lastpage}
\end{document}